\documentclass[12pt]{iopart}
\usepackage{epsfig,graphicx}

\usepackage{iopams}  

\newcommand{\bl}[1]{\begin{equation}\label{#1}}
\newcommand{\be}{\begin{equation}}
\newcommand{\ee}{\end{equation}}
\newcommand{\bea}{\begin{eqnarray}}
\newcommand{\eea}{\end{eqnarray}}

\newcommand{\td}[2]{\frac{\mathrm{d}{#1}}{\mathrm{d}{#2}}}
\newcommand{\rec}[1]{\frac{1}{#1}}
\newcommand{\z}[1]{\left({#1}\right)}
\newcommand{\sz}[1]{\left[{#1}\right]}
\newcommand{\kz}[1]{\left\{{#1}\right\}}

\renewcommand{\v}[1]{\mathbf{#1}}

\renewcommand{\r}[1]{(\ref{#1})}
\newcommand{\eq}[1]{eq.~(\ref{#1})}
\newcommand{\eqs}[2]{eqs.~(\ref{#1}) and (\ref{#2})}

\begin{document}

\title[New exact solutions of relativistic hydrodynamics]
      {New exact solutions of relativistic hydrodynamics}

\author{T.~Cs\"org\H{o}$^1$,  M.~I.~Nagy$^1$, and M.~Csan\'ad$^2$}

\address{$^1$ MTA KFKI RMKI, Budapest, H - 1525, Hungary \\
         $^2$ ELTE, E{\"o}tv{\"o}s Lor{\'a}nd University, H - 1117 Budapest, P{\'a}zm{\'a}ny P. s. 1/A, Hungary}
\ead{csorgo@rmki.kfki.hu}%

\begin{abstract}
A new class of simple and exact solutions of relativistic hydrodynamics 
is presented, and the consequences are explored in data analysis.
The effects of longitudinal work and acceleration are taken into account in an
advanced estimate of the initial energy density  and the life-time of the 
reaction. 
\end{abstract}


\section{Introduction: High temperature superfluidity}
Fluid dynamics is based only on local conservation laws and thermodynamics, 
so it can be successfully applied
to a vast range of physical phenomena. Recently,
a new state of matter has been created in Au+Au collisions at RHIC,
and, surprisingly, it was found to flow as a perfect fluid~\cite{PHENIX-White}.
The kinematic shear viscosity of this nearly perfect fluid has been 
determined and found to be at least a factor of 4 smaller, 
than that of the superfluid $^4$He ~\cite{Lacey:2006bc}.
Let us refer to this property as {\it high temperature superfluidity}, 
to highlight that the fluid of quarks in the super-high, $T \simeq 2$ terakelvin temperature range
flows better than the superfluid $^4$He in the extremely low, 1-4 K temperature region.

Exact solutions of perfect fluid hydrodynamics have powerful implications in  
high energy particle and nuclear physics. 
Although the renowned Landau-Khalatnikov 
solution~\cite{Landau:1953gs,Khalatnikov,Belenkij:1956cd} is only implicitly given,
it describes well the energy dependent increase of the width of the rapidity distribution. 
Although the renowned Hwa-Bjorken solution~\cite{Hwa:1974gn} 
lacks acceleration and yields a too perfectly flat rapidity distribution,
 it yields a key estimate of the initial energy density in high energy reactions. 
Other families of exact solutions were born
from the desire of understanding 
the dynamics of high-energy heavy reactions~\cite{Chiu:1975hw,Baym:1984sr,Srivastava:1992cg,Eskola:1997hz,Biro:2000nj,Csorgo:2003rt,Csorgo:2003ry,Sinyukov:2004am}. 
Here we present a new, accelerating, analytic,  exact and explicit solution of 
relativistic hydrodynamics, which fits $dn/dy$ data at RHIC and allows for an 
advanced estimate of initial energy density and life-time of the reaction. 

\section{Simple solutions of relativistic hydrodynamics}
We discuss spherical (and one-dimensional) solutions, where $r$ denotes the radial
 (or the single spatial) coordinate, and $d$ is the number of dimensions. 
The pressure is denoted by $p$\,, the energy density by $\varepsilon$\,, and the
temperature by  $T$,
the four-velocity by $u^{\mu}=\gamma(1,\v{v})$, 
and $\v{v}=v\v{n}$ is the three-velocity. 
 For the equation of state (EoS), we assume that  
$\varepsilon  - B =\kappa (p+B) $, with $\kappa=1/c_s^2$, where $c_s$ 
is the speed of sound and $B$ stands for a bag constant (that may have a vanishing value 
too). In high energy collisions, the entropy density 
$\sigma$ is large, but the net charge density is small. Thus 
we assume that all the conserved charges have zero chemical potential. 
In perfect fluids, $\sigma$ and four-momentum tensor $T^\mu_\nu=(\varepsilon + p) u^\mu u_\nu-p\delta^\mu_\nu$ are locally conserved:
$\partial_\nu(\sigma u^\nu)=0$\,, $\partial_\nu T^{\mu\nu}=0$. With projections, we obtain the
Euler and the energy equations:
\bea
(\varepsilon+p)u^{\nu}\partial_{\nu}u^{\mu}&=& \z{\delta^\mu_\rho-u^\mu u_\rho}\partial^\rho p, \label{Reul} \\
(\varepsilon + p)\partial_{\nu}u^{\nu}+u^{\nu}\partial_{\nu}\varepsilon &=& 0. \label{RE}
\eea
Our solutions can be written down in the so-called Rindler-coordinates $\tau$ and $\eta$: $t=\tau\cosh\eta$,
$r=\tau\sinh\eta$. We found the following expression for $p$ and $v$:
\bl{e:sol} v\, =\, \tanh\,\lambda\eta \,,\quad p+B \, = \, p_0\z{\frac{\tau_0}{\tau}}^{\lambda
d\frac{\kappa+1}{\kappa}}\z{\cosh\frac{\eta}{2}}^{-(d-1)\phi_{\lambda}} . 
\ee
We obtain solutions in some special cases of the constants listed in Table I, and for any $B$. In what follows, we set $B=0$ to simplify the description.
\begin{table}
\begin{center}
\begin{tabular}{|c|c|c|c|c|}
  \hline
  \hline
 Case: & $\lambda$ & $d$ & $\kappa$ & $\phi_{\lambda}$ \\
  \hline
  \hline
 (a)  & $2$             & $\in\mathbb{R}$ & $d$              & $0$ \\ \hline
 (b)  & $\rec{2}$       & $\in\mathbb{R}$ & $1$              & $\frac{\kappa + 1}{\kappa}$ \\ \hline
 (c)  & $\frac{3}{2}$   & $\in\mathbb{R}$ & $\frac{4d-1}{3}$ & $\frac{\kappa + 1}{\kappa}$ \\ \hline
 (d)  & $1$             & $\in\mathbb{R}$ & $\in\mathbb{R}$  & $0$ \\ \hline
 (e)  & $\in\mathbb{R}$ & $1$             & $1$              & $0$ \\ \hline
  \hline
\end{tabular}
  \caption{New exact hydrodynamical solutions are given by lines (a, b, c) and (e), while
case (d) is the already known Hwa-Bjorken (and Hubble) solution.}\label{t:1}
\end{center}
\end{table}
While case (d) is the Hwa-Bjorken solution, case (a) is a new solution valid in arbitrary dimensions.
It has uniformly accelerating trajectories. Case (b) and (c) were found jointly by
T.~S.~Bir\'o~\cite{Biro:2007pr} (for $d=1$) and by us (for any $d$). In these solutions $p$ is finite in $\eta$. Case (e) has a special EoS and is
valid only in 1+1 dimensions, but $\lambda$ can be any real number. This is why we 
shall use it in the following applications. 

\section{Applications to high-energy reactions} 
\begin{figure}[tb]
\begin{center}
 \includegraphics[width=200pt]{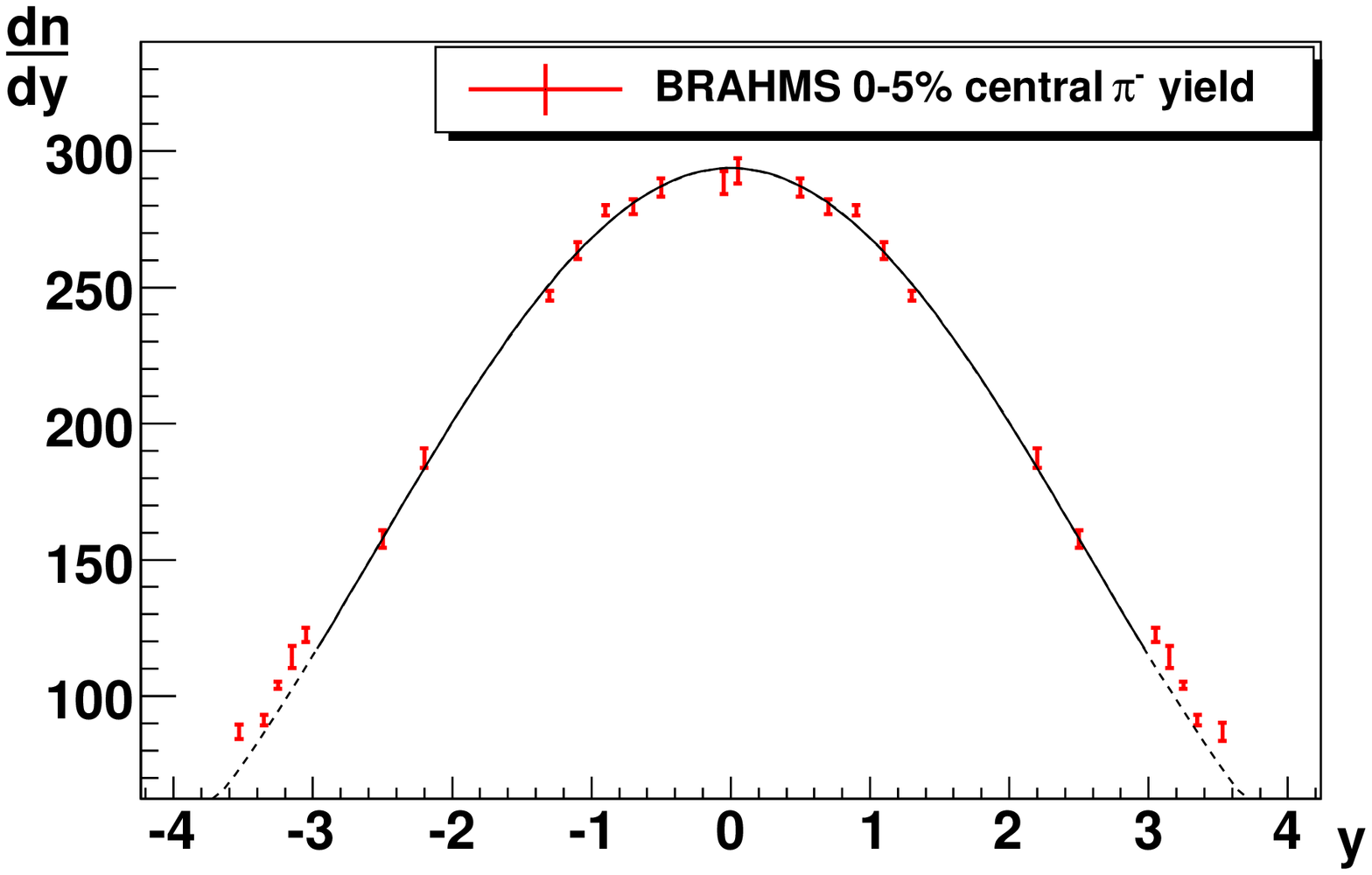} 
 \includegraphics[width=200pt]{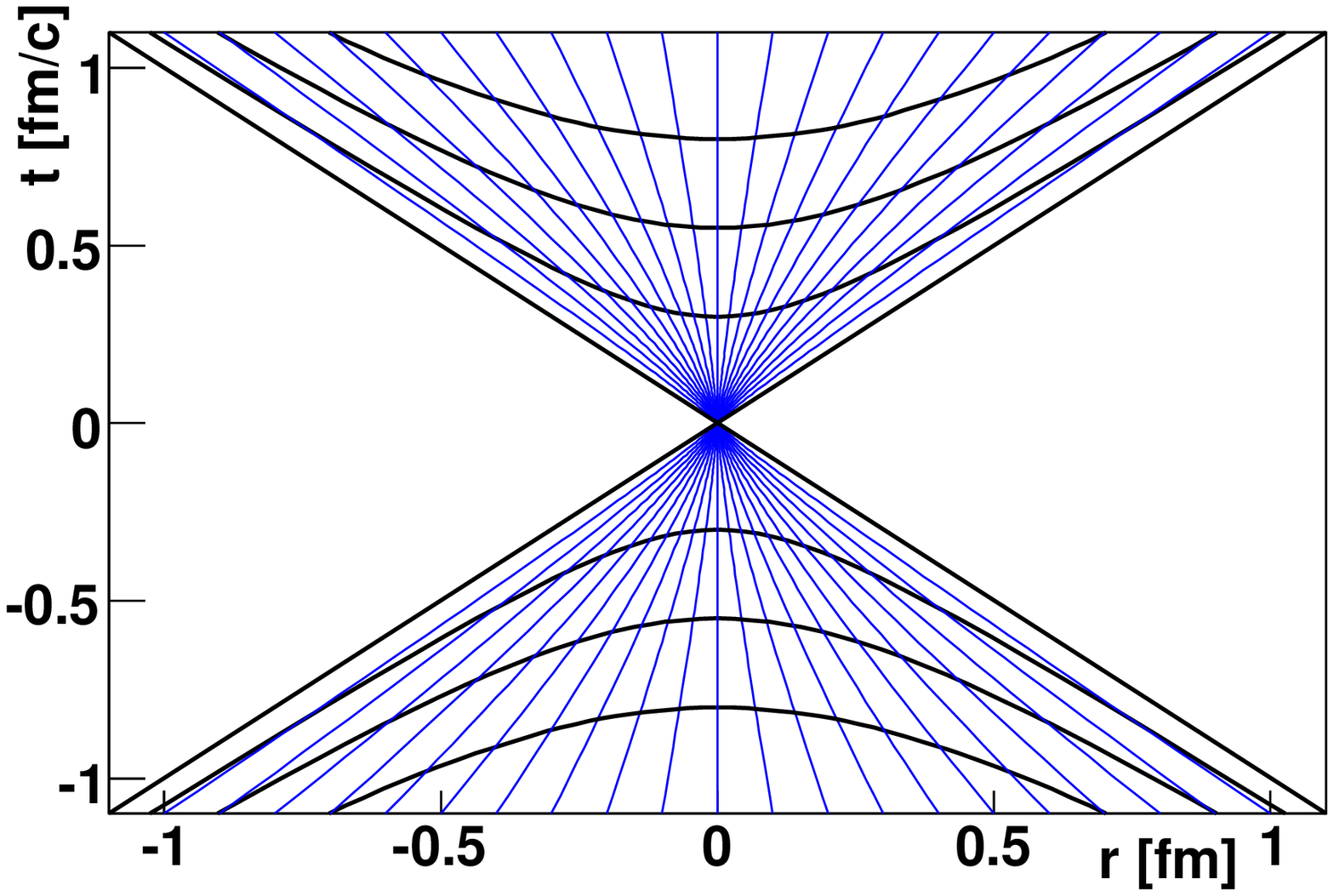} 
 \caption{Left panel: Result of the fit of \eq{e:dndy-approx} to BRAHMS $\td{n}{\eta}$ data of
          Ref.~\cite{Bearden:2004yx}. Right panel: Flow trajectories and possible freeze-out
          hypersurfaces of the $\lambda=1.18$ solution, which fits to BRAHMS $\td{n}{y}$ 
	data. (The physical solution is in the future light cone, but for
	aesthetic reasons its  extension to the past light cone is also shown).}\label{f:dndeta}
\end{center}
\end{figure}

\begin{figure}
\begin{center}
 \includegraphics[width=200pt]{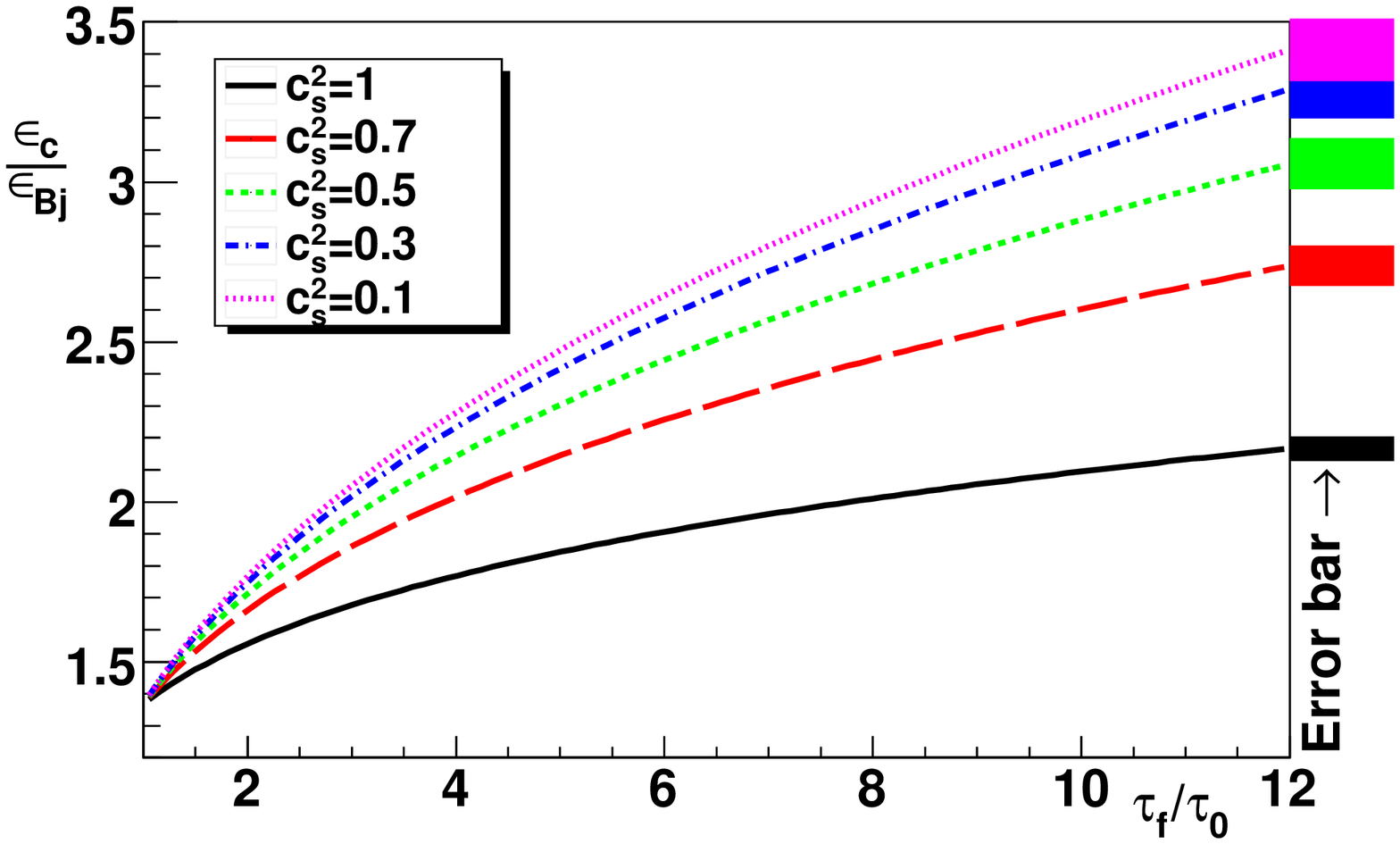}
 \includegraphics[width=200pt]{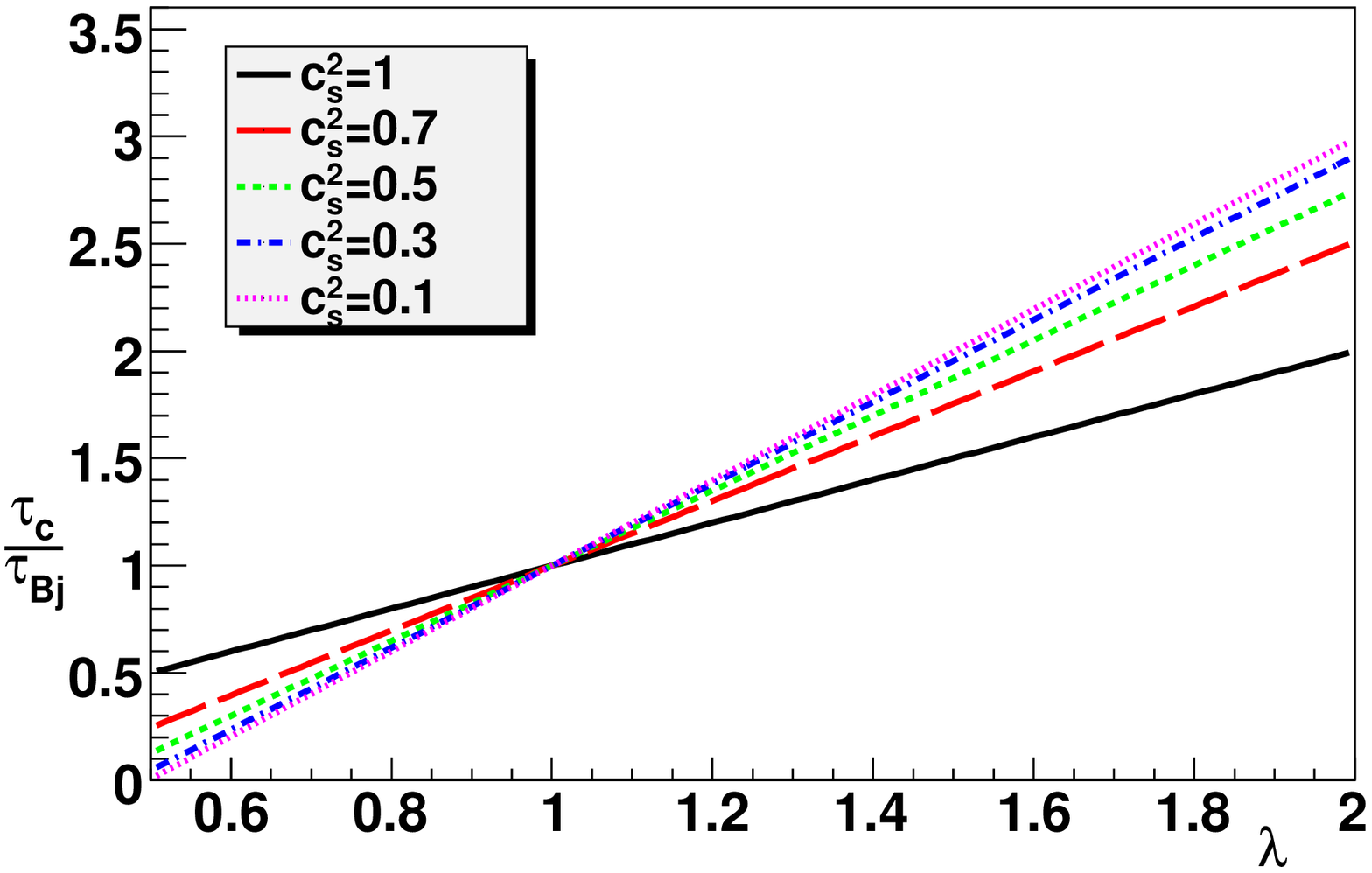}
 \caption{Dependence of $\varepsilon_{c_s^2}/\varepsilon_{Bj}$ 
on $\tau_f/\tau_0$  (left panel) and $\tau_{c_s^2}/\tau_{Bj}$
on $\lambda$   (right panel) for various $c_s^2$ values. Solid lines for $c_s^2=1$ indicate the
          exact result of \eqs{e:ncscs}{e:Rlong-c}, the others show the conjectures of \eq{e:conj}.}\label{f:tau}
\end{center}
\end{figure}

For the description of high-energy reactions, we calculated the rapidity distribution,
$\td{n}{y}$ from the solution denoted as case (e) in Table I. Assuming sudden freeze-out
at temperature $T(\tau_f,\eta=0)=T_f$ on a hypersurface pseudo-orthogonal to $u^\mu(x)$, with a saddle-point
integration in $\eta$ and the transverse mass $m_T$ (which becomes exact if $m/T_f \gg 1$, where $m$ is the
particle mass), we got the following formula~\cite{detailed}:
\bl{e:dndy-approx}
\td{n}{y} \approx \td{n}{y}\Big{|}_{y=0}
   \cosh^{-\frac{\alpha}{2}-1}\z{\frac{y}{\alpha}}
   \exp\kz{-\frac{m}{T_f}\sz{\cosh^\alpha\z{\frac{y}{\alpha}}-1}} ,
\ee
with $\alpha=\frac{2\lambda-1}{\lambda-1}$. 
The parameter $\Delta y^2 = \frac{\alpha }{m/T_f + 1/2 + 1/\alpha}$ characterizes this distribution: it
has a  minimum at $y=0$, if $\Delta y^2<0$, it is flat if $\Delta y^2=0$, (this is the case when
$\lambda = 1$), otherwise it is nearly Gaussian. We extracted the $\lambda$ parameter for collisions
at $\sqrt{s_{NN}}=200$ GeV by fitting \eq{e:dndy-approx} to $\td{n}{y}$ data measured by the
BRAHMS collaboration~\cite{Bearden:2004yx}. Fig.~\ref{f:dndeta} shows the fit, and the corresponding flow profile.
We found $\lambda=1.18\pm 0.01$, which indicates the presence
of acceleration, as shown in Fig 1 by the curvature of the fluid lines. 
This influences both the estimation of initial energy density (because of the faster
expansion and the work done by the matter), and the estimation of the life-time of the reaction.
This way we found a new estimate of the initial energy density $\varepsilon_c$ 
that generalizes Bjorken's estimation, $\varepsilon_{Bj}$ to non-flat rapidity distributions:
\bl{e:ncscs}
\frac{\varepsilon_c}{\varepsilon_{Bj}}=\z{2\lambda-1}\z{\frac{\tau_f}{\tau_0}}^{\lambda-1}\,\qquad \mbox{where} \quad
\varepsilon_{Bj}=\frac{\langle m_t\rangle}{(R^2\pi)\tau_0}\frac{dn}{dy}\Big{|}_{y=0} .
\ee
The life-time estimation from the longitudinal length of homogeneity, $R_{{\rm long}}$ of Sinyukov and
Makhlin, $\tau_{Bj}$ was also based on the Hwa-Bjorken solution~\cite{Makhlin:1987gm}. Our new estimation,
$\tau_{c}$, which takes acceleration effects into account, is 
\bl{e:Rlong-c}
R_{{\rm long}}=\sqrt{\frac{T_f}{m_t}}\frac{\tau_{c}}{\lambda} \qquad\mbox{thus}
	\quad \tau_{c} = \lambda \tau_{Bj} =
	\lambda \sqrt{\frac{m_t}{T_f}}R_{{\rm long}}
 .
\ee
The \eqs{e:ncscs}{e:Rlong-c} show that both the initial energy density and the life-time of the reaction 
is under-estimated by formulas based on the accelerationless of Hwa-Bjorken solution. 
For more physical EoS than that super-hard $\kappa = 1$ corresponding to the $\lambda=1.18$
solution, we conjectured new, advanced estimations of the initial energy density  $\varepsilon_{c_s^2}$ and
the life-time of the reaction $\tau_{c_s^2}$ as
\bea
\frac{\varepsilon_{c_s^2}}{\varepsilon_{Bj}} & = &\z{2\lambda-1}\z{\frac{\tau_f}{\tau_0}}^{\lambda-1}
\z{\frac{\tau_f}{\tau_0} }^{(\lambda-1)(1-c_s^2)}, \label{e:conj} \\
\tau_{c_s^2} & = &[\lambda + (\lambda -1)(1 - c_s^2)] \tau_{Bj} .
\eea
 For their detailed explanation, see Refs.~\cite{Csorgo:2006ax,detailed}.
 Using the Bjorken estimate of $\varepsilon_{Bj} = 5$
GeV/fm$^3$ as given in Ref.~\cite{BRAHMS-White}, and $\tau_f/\tau_0=8\pm 2$ fm/c, our analytic
formula \r{e:ncscs} yields $\varepsilon_c = (2.0\pm0.1)\varepsilon_{Bj}=10.0\pm 0.5$ GeV/fm$^3$
for $\lambda=1.18\pm 0.01$. For the life-time, \eq{e:Rlong-c} implies a 18$\pm$1\% increase.
Fig.~\ref{f:tau} shows the speed of sound dependence of $\varepsilon_{c_s^2}/\varepsilon_{Bj}$ as a function of $\tau_f/\tau_0$, and that of
$\tau_{c_s^2}/\tau_{Bj}$ as a function of
$\lambda$.
For a realistic equation of state, where $c_s^2 = 0.1$
in  $\sqrt{s_{NN}}= 200$ GeV Au+Au reactions~\cite{Adare:2006ti}, 
acceleration increases the initial energy density estimate to 
15 GeV/fm$^3$, and the  life-time estimate by 36 \%.

\bigskip
\end{document}